**Superstrong nature of covalently bonded glass-forming liquids at select compositions**

K. Gunasekera[1], S. Bhosle[1], P. Boolchand[1] and M. Micoulaut[2]

[1] *School of Electronics and Computing systems, College of Engineering and Applied Science, University of Cincinnati, Cincinnati, OH 45221-0030*

[2] *Laboratoire de Physique Théorique de la Matière Condensée, Université Pierre et Marie Curie, Boite 121, 4, Place Jussieu, 75252 Paris Cedex 05, France.*

Variation of fragility (m) of specially homogenized $Ge_xSe_{100-x}$ melts are established from complex specific heat measurements, and show m(x) has a global minimum at an extremely low value (m=14.8(0.5)) in the 21.5% < x < 23% range of Ge. Outside of that compositional range, m(x) then increases at first rapidly then slowly to about m=25-30. By directly mapping melt stoichiometry as a function of reaction time at a fixed temperature T>$T_g$, we observe a slowdown of melt-homogenization by the super-strong melt compositions, 21.5% < x < 23%. This range furthermore appears to be correlated to the one observed between the flexible and stressed rigid phase in network glasses. These spectacular features underscore the crucial role played by topology and rigidity in the properties of network-forming liquids and glasses which are highlighted when fragility is represented as a function of variables tracking the effect of rigidity. Finally, we investigate the fragility-glass transition temperature relationship, and find that reported scaling laws do not apply in the flexible phase, while being valid for intermediate and stressed rigid compositions.

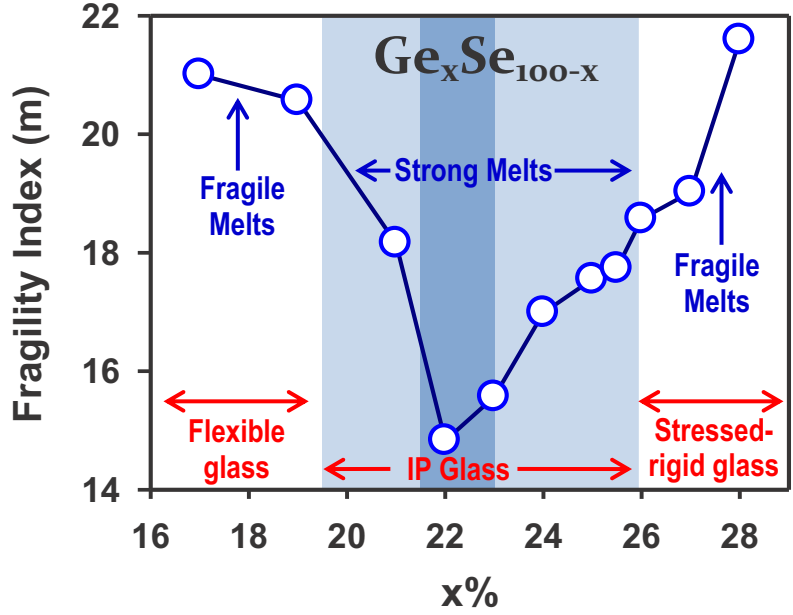

TOC Figure

I. INTRODUCTION

The strong-fragile classification of supercooled liquids based on the temperature evolution of dynamic quantities[1-4] such as viscosity ($\eta$) has proved useful in understanding viscous slow down as melts are cooled to $T_g$, the glass transition temperature. However, in network forming liquids the connection between fragility, thermodynamics, network rigidity and topology remains elusive[5-7]. Both viscosity and structural relaxation time ($\tau \propto \eta$) increase enormously as T is lowered to $T_g$, the temperature at which $\eta$ acquires an astronomically high value of $10^{12}$ Pa.sec, while $\tau$ increases to a characteristic value of about 100 seconds. A useful way to characterize how $\tau$ (or $\eta$) approaches the glass transition is provided[4] by the slope of log($\tau$) with temperature, near $T_g$. This gives the fragility index as:

$$m = \left[\frac{d\log(\tau)}{dT_g/T}\right]_{T \to T_g} \tag{1}$$

As most of the glass-forming liquids display an Arrhenius behavior when T→$T_g$, the slope characterizing the variation of log($\tau$ ) with 1/T can be used to define a corresponding (apparent) activation energy given by: $E_a$ = m.$T_g$.ln(10). Experiments on a wide variety of supercooled liquids[4] reveal that fragility typically can vary over a wide range[8], 15 < m < 175, with the lowest value characteristic of *strong liquids* possessing an $E_a$ that is T-independent, while the higher values of (m) identified with *fragile liquids* that show an $E_a$ that steadily decreases with 1/T, and leads to an increasingly non-exponential variation of $\tau$ or $\eta$. The latter is usually described in terms of the Vogel-Fulcher-Tammann (VFT) function of the form exp[*A/(T − T₀* )], where *T₀* < *$T_g$* is a characteristic temperature at which dynamics diverge, and A a fitting parameter. Note that as m increases, the departure from Arrhenius variation or non-exponentiality sets in[4] given the similar limit of viscosity at high temperature ($10^{-3}$-$10^{-4}$ Pa.s) for both, strong or fragile[9] liquids. But there is an important difference in dynamics between fragile and strong glass-forming melts: it is manifested in a lower diffusivity D (D ∝ 1/$\eta$ )[10] in the latter (strong) at T > $T_g$ but that trend reverses at T < $T_g$[11]. Given the fact that fragility can be tuned with chemical composition[12], one may observe important effects on melt homogenization as starting materials are alloyed. On a microscopic scale, local compositions, different from the nominal one, will appear and possess different fragility and diffusivity. At a macroscopic scale, it is well known that chalcogenide supercooled melts display a fragility minimum at certain compositions[13,14] identified with a flexible to rigid transition[15]. However, a more general correlation between homogenized liquids displaying no phase-separation (T > $T_g$) and corresponding glasses (T < $T_g$ ) upon viscous slow down and the onset of rigidity and stress transitions has never been established[16].

Can a clear relationship be drawn between the fragility of a glass-forming liquid and the ease or difficulty of homogenization of the corresponding melt? The present study attempts to answer this basic question by connecting compositional changes in network topology with the viscous slow down occurring close to the glass transition.

In this article we show that *super-strong* network glass-forming liquids (m=14.8(0.5)) existing in a narrow compositional range act as barriers to the process of melt homogenization due to their high viscosity. These barriers are intimately related to the underlying topology and rigidity of the network structure. We base our conclusions from measurements of fragility on the specially homogenized $Ge_xSe_{100-x}$ melts [17-19] in the 10% < x < 33.33% range using complex specific heat, $C_p(\omega,T)$ measurements. Modulated DSC permits [20,21] extending traditional relaxation studies (dielectric relaxation, viscosity) at high frequency to extremely low frequencies of 0.06 sec$^{-1}$, affording fragility measurements close to $T_g$. This has the advantage that fragility measurements can be extended to super cooled melts that easily crystallize at $T > T_g$, where viscosity measurements are not feasible. Such is the case in the present binary where melts exceeding Ge content of 27 mole% easily crystallize above $T_g$. Dielectric measurements in the context of glass transition can also be extended to low frequency [22], and fragility index measurements from calorimetric spectroscopy appear to be fully consistent[20,21] with those reported from dielectric data. The enthalpy relaxation time ($\tau_e$) of melts near $T_g$ were measured, and melt fragility (m) established using equation (1) and then activation energy $E_a$ from m. Our results reveal that $Ge_xSe_{100-x}$ melts in the narrow composition range, 21.5% < x < 23%, possess a very low fragility m=14.8(0.5), lower than the well-known silica example[4], i.e. display *super-strong* behavior. The fragility of the archetypal fused $SiO_2$ reported by several groups is in the 20 to 28 range[23], and appears to be uncertain because of impurities that influence $T_g$ and the dynamics themselves[24]. Nevertheless, fused $SiO_2$ is widely viewed as a strong liquid with a fragility greater than the usual reported value of 16 for strong glass-forming liquids[4]. Raman scattering acquired along length of melt columns, as starting materials are reacted at elevated T, has permitted us to directly map the evolution of "melt stoichiometry". We find that slow melt homogenization of these covalently bonded networks can be traced to presence of '*super-strong*' melt inclusions that serve as a bottleneck in melt-mixing.

## II. EXPERIMENTAL

### A. Synthesis and Raman profiling

Specially homogenized bulk $Ge_xSe_{100-x}$ glasses were synthesized by reacting the 99.999% elemental Ge and Se pieces from Alfa Aesar and vacuum ($10^{-7}$ Torr) sealed in 5mm ID fused quartz tubes. Care was taken to work with 3 to 4 mm size pieces of the elements and kept dry. Two gram sized batches held vertically were reacted at 950°C in a T- programmable box furnace for an extended period, $t_R$, ranging up to 9 days. Periodically samples were water quenched and FT- Raman spectra accumulated at 10 locations 2.5 mm apart on the 25 mm length melt column encased in quartz tubes. The laser spot size was kept at 50 μm. Prior to quenching, melts were equilibrated for 30 minutes 50° C above the liquidus. Fig. 1 illustrates results obtained at x = 23% after reaction times, $t_R$, indicated in the 8 panels. Note that after a short time $t_R$ = 6h (Fig.1a), one observes crystalline phases (narrow Raman bands) to form at the tube bottom, but with continued reaction these phases dissipate and Raman spectra characteristic of glasses appear at $t_R$ > 24h. It is useful to mention that even though we did not rock the samples, in the early stages of melt reaction, liquid Se vigorously runs up and down the melt column since the reaction temperature (950°C) far exceeds the Se melting point (220.8°C) but not that of Ge melting point (937.4 °C). Molten Ge formed at the tube bottom reacts with the flowing Se, forming Ge-rich crystalline- and amorphous-phases. With continued reaction, $t_R$ = 24h, these phases dissipate and bulk glasses of variable stoichiometry form along the length column (Fig1b). And as melts are reacted longer, it is only after $t_R$ = 216h (Fig. 1h), that the 10 lineshapes taken along the melt column became identical, providing the clearest signature that the batch as a whole has homogenized. Fig.2 illustrates results obtained for a melt weighed at x = 21%. The results are qualitatively similar except the melt homogenized in 144h (Fig. 2d), i.e., in a shorter time than the 219h required to homogenize a melt at x =23% (Fig.1h).

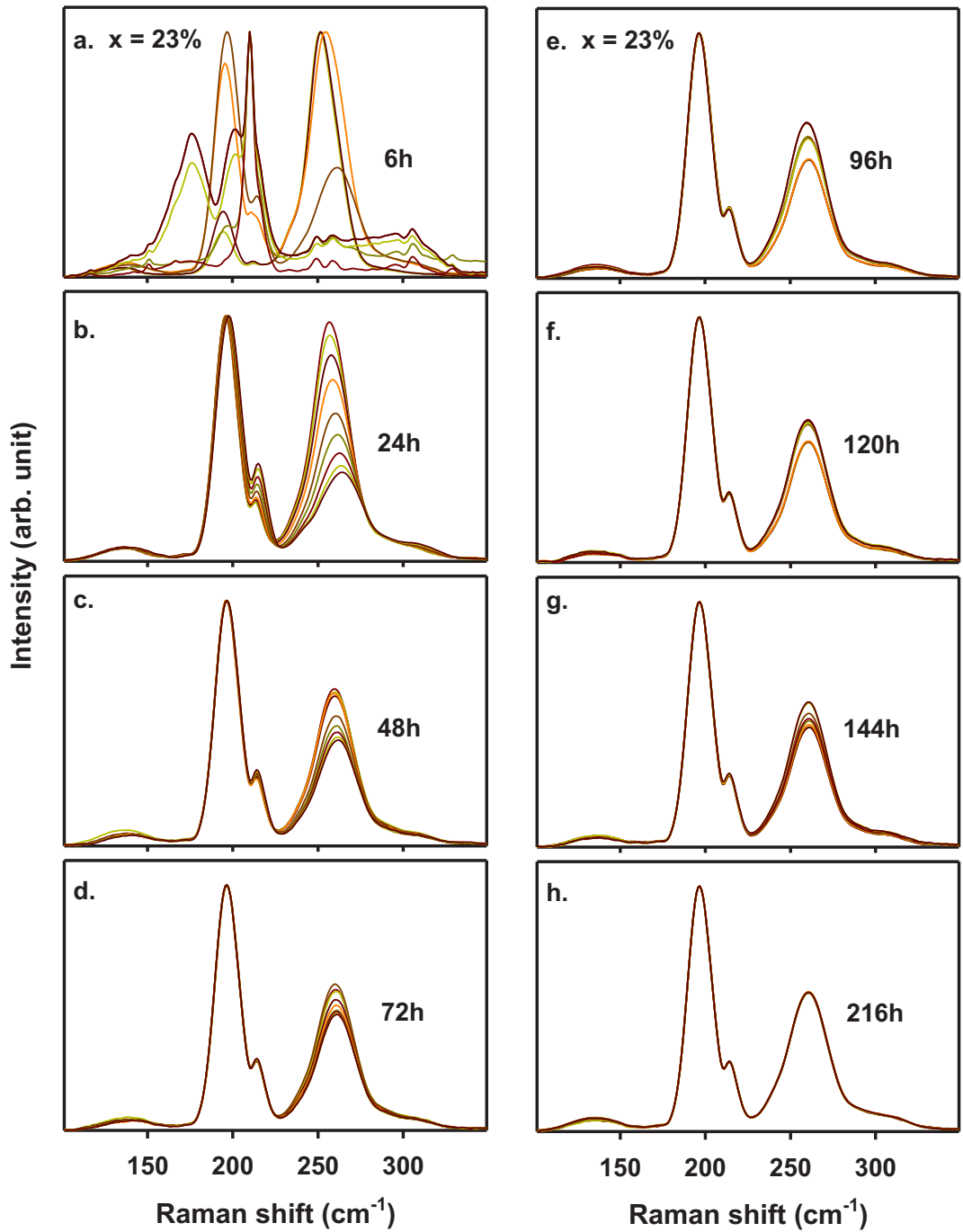

**FIGURE 1:** FT- Raman spectra taken at 9 locations along the length of a melt-quenched column at $Ge_{23}Se_{77}$. As melts are reacted at 950°C for steadily longer times $t_R$, the spread in the lineshapes decreases, and eventually vanishes at $t_R = 216h$, signaling that the 2 gram batch has completely homogenized.

To gain a basic understanding how melts homogenize, we then proceeded to analyze the observed lineshapes and have extracted the scattering strength ratio of the Se-chain mode near 250 cm$^{-1}$ to the GeSe$_4$ Corner Sharing (CS) mode near 200 cm$^{-1}$. The details of least-squares fitting the observed lineshapes are provided elsewhere[17,18]. Prior to this work we had at our disposal a library of the Raman scattering strength ratios for various modes[17-19] for the completely homogenized Ge$_x$Se$_{100-x}$ melts/glasses at every 2 mole% increment of x. Using the library, we deduced the melt stoichiometry 'x' at a given height h along a quartz tube. Fig.3 summarizes the h(x) data for a melt weighed at x = 23%. After $t_R$ = 24h (1d) melts display a variation in stoichiometry along the length from nearly x = 28% at the lowest point (h= 1) to about x = 16% at the highest point (h = 9). This behavior is as expected given that the liquid Ge density of 5.60 gms/cm$^3$ exceeds that of the liquid Se of 3.99 gms/cm$^3$. It is for these reasons that in the very early stages ( Fig.1a, $t_R$ = 6h), Ge-rich crystalline and amorphous phases are formed[17,18] at the tube bottom as we alluded to earlier.

The smooth variation of melt stoichiometry during such a synthesis process is a strong asset. We could reliably ascertain the melt stoichiometry variation even though we sampled only 2% of the length column (10 spots of 50 μm in size versus 25 mm) in the Raman profiling experiments. Such is not the case if one rocks the samples, as off-stoichiometric inclusions formed are distributed randomly and hard to detect with a 2% sampling. The manner in which melt homogenize is fascinating and we discuss the underlying issues in section III.

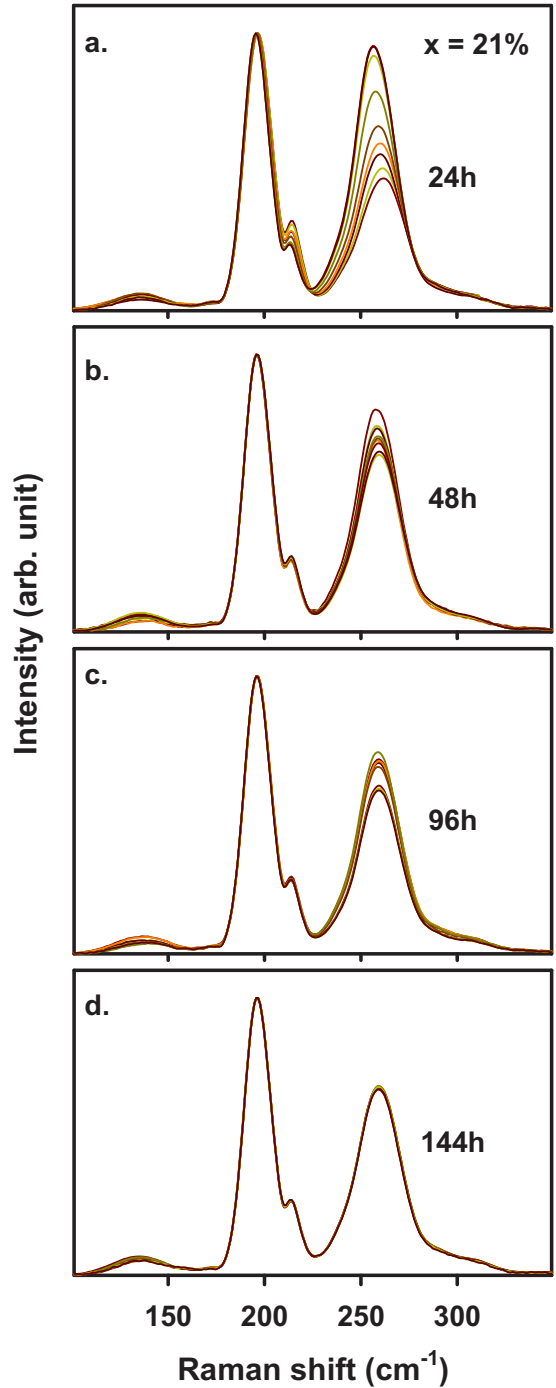

**FIGURE 2:** FT- Raman profiles of a melt at $Ge_{21}Se_{79}$ examined as a function of reaction time of the starting materials at 950°C, reveals the batch to completely homogenize at $t_R$ = 144h. The kinetics of melt homogenization for $Ge_{23}Se_{77}$ melt shown in Fig.1 are considerably slower because of its super-strong nature or low fragility.

### B. Fragility

We used a TA instrument model Q2000 unit to examine the glass transition endotherm in terms of complex $C_p$ formalism as illustrated in Fig 4a for the case of a melt at x = 10%. The imaginary part of $C_p$ ($C_p''$) (also related to the non-reversing heat flow) shows a peak when $\omega\tau_e$ = 1, i.e., when the inverse of the melt enthalpic relaxation time ($\tau_e$) tracks the modulation frequency ($\omega$)[20]. But as $\omega$ increases, the step in real part of $C_p$ ($C_p'$) (also related to the reversing heat flow), and the peak in $C_p''$ shifts to higher T as expected. One defines $T_g$ by the peak in $C_p''$ when $\omega = 2\pi/(t_{mod}) = 0.06$ sec$^{-1}$ or $t_{mod}$ = 100 sec. In earlier work[17,18] we measured $T_g$ by the inflexion point of the reversing heat flow associated with the glass transition endotherm measured at a modulation time period of $t_{mod}$ = 100 sec . Those glass transitions are found to be *identical* to the present ones deduced above from the peak of $C_p''$.

Our results show (Fig. 5) that as x increases from pure Se (x = 0) the activation energy steadily decreases at first slowly but then sharply near x = 22% to acquire a global minimum of 139(5) kcal/mol. At higher x > 25% , $E_a(x)$ then increases rather rapidly largely reflecting the $T_g(x)$ increase through the relation

$$E_A = m \cdot T_g \cdot \ln(10). \qquad (2)$$

Melt fragility were accessed directly from the variation of $\tau(T)$ in Fig. 4. Compositional trends of melt fragility over an extended range, m(x), show (Fig.6) a global minimum of m = 14.8(0.5) in the 21.5% < x < 23% range, a result that confirms the fragility minimum reported earlier[14,25] near x = 22.5% from viscosity measurements at higher temperatures. The fragility at x = 0, i.e., for pure Se was also measured and found to be 51.1(0.5) indicating that a glass made of Se$_n$ chains is fragile. For completeness we have included in Fig. 6b, the T$_g$ of the present

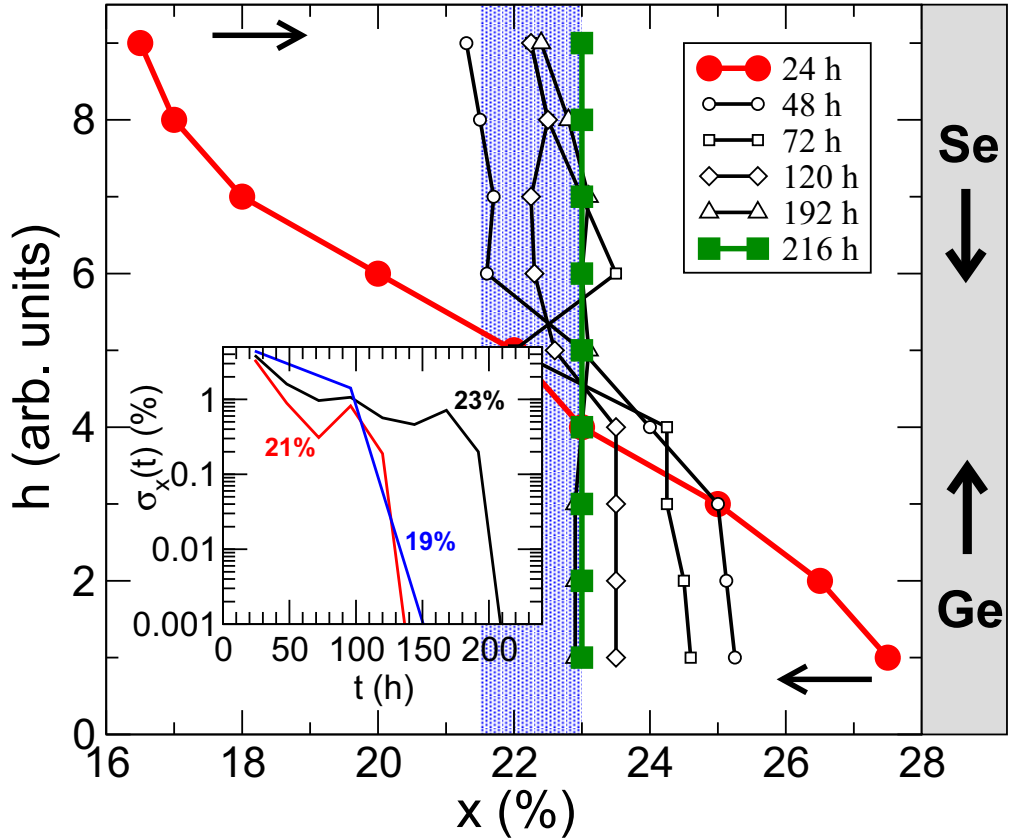

**FIGURE 3:** Data of Fig.1 is used to map melt stoichiometry variation as a function of t$_R$, at x = 23%. Here h represents the height of melt column in units of 2.5 mm from the tube bottom (h = 0), and x the Ge content of the melt deduced from the Raman spectrum. Note that at t$_R$> 3days the melt-mixing kinetics are arrested in the top half (5< h< 9) of the column, as it negotiates through the super-strong compositions (shaded area). In the lower half, (1< h< 4), the kinetics of melt-mixing do not see that arrest since melt compositions always reside outside the super-strong compositions. The inset shows the standard deviation of compositions σ$_x$(t) with time for selected compositions (see text for details). Right: schematic view of the diffusion of the reacting elements along the melt column.

glasses measured in MDSC using a modulation time period $t_{mod}$ = 100 sec. The kinetic shifts associated with these $T_g$ due to finite scan rates were eliminated by recording a cooling scan following a heating one as discussed elsewhere[17,18]. The trend reveals a monotonic increase of

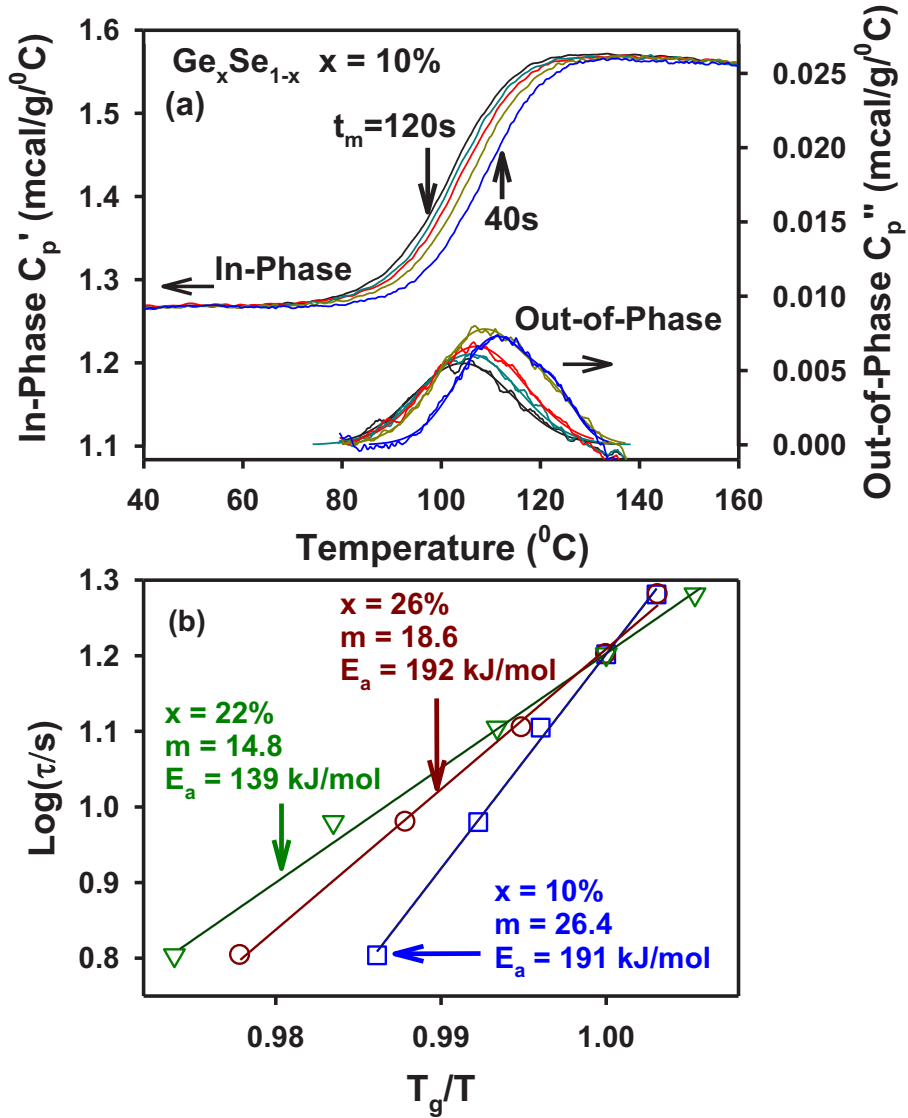

**FIGURE 4:** (a) *In-phase* and *out-of-phase* components of complex $C_p$ from modulated-DSC scans as a function of modulation frequency for a $Ge_xSe_{100-x}$ melt at x = 10%. (b) Log of relaxation time ($\tau_e$) plotted as a function of $T_g/T$ yielding fragility, m, and activation energy $E_a$ from the slope of the Arrhenius plots.

$T_g(x)$ across the 19.5% < x < 26% range, wherein a square-well like minimum[19] in the non-reversing enthalpy of relaxation at $T_g$ is manifested (lightly shaded region of Fig.6a), also known as the reversibility window[19] in glasses (T < $T_g$), and also a global minimum of m(x) in corresponding melts (T > $T_g$) (darkly shaded region in Fig.6a) as found in the present work. We shall discuss these results next.

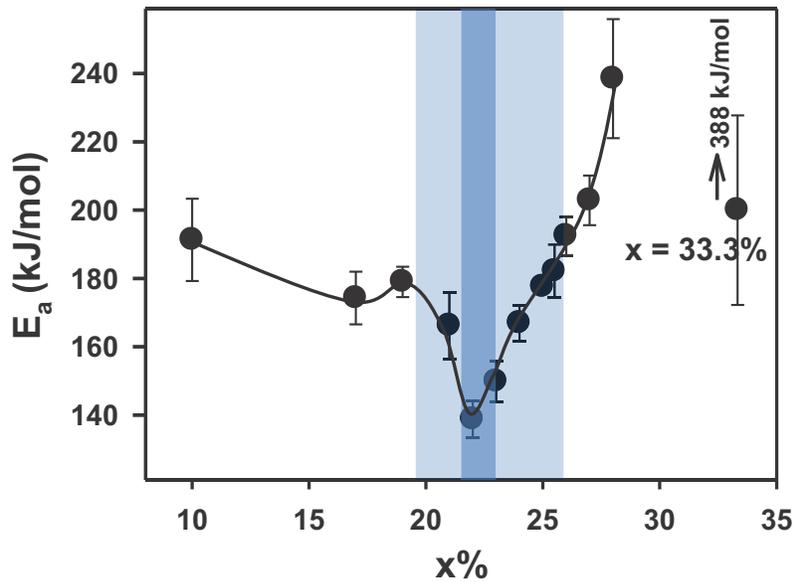

**FIGURE 5:** Variation of activation energy $E_a$ (x) from present calorimetric measurements. Note the sharp lowering of $E_a$ for melt compositions in the super-strong region (dark blue). The light blue band gives the intermediate phase of the present system[18].

### III. DISCUSSION

There are several central issues raised by the present work. First, why do Ge-Se melts homogenize so slowly? We will show the result is closely connected to the super-strong nature of select melt compositions. Second, in the compositional range where networks undergo elastic phase transitions, such as from flexible phase to rigid phase, melts ( T > $T_g$ ) display a global minimum in fragility, while corresponding glasses ( $T < T_g$) show a global minimum in the non-reversing enthalpy of relaxation at $T_g$. These new findings, observed in the homogeneously synthesized melts/glasses of the present binary, apparently are not peculiar to the present binary but appear to be a generic feature observed in many other glass systems. We discuss the broad consequences of these observations as well.

### A. Super-strong character of melts and kinetics of homogenization

Our choice of the composition x = 23% in the $Ge_xSe_{100-x}$ binary for mapping the kinetics of melt homogenization (fig.3) using Raman profiling is based on two factors. First the composition lies at the center of the Intermediate Phase (19.5% < x < 26.0%) of corresponding glasses[19], and second, present fragility results show such melts to be rather strong, i.e., display close to a minimum of $E_a$ = 139(5) kJ/mol (Fig. 5) and a minimum of m = 14.8(0.5) (Fig.6a). For such a melt at x = 23% , at $t_R$ = 24 h (Fig.3), in the initial stages, we have already alluded to the fact that lower half of the batch column (h < 5) is Ge-richer than the top half (h > 5), a reflection of the higher densities of melts in the lower half compared to the top half. The process of homogenization entails Ge atoms diffusing up and Se moving down a melt column as schematically illustrated in the right panel of Fig.3. In the early phase ($t_R$ < 3 days) of the reaction process as melt densities nearly equalize, diffusion (D) of Se atoms down the melt column *exceeds* that of Ge atoms up the column, largely because  D ~ $1/(\rho)^{1/2}$. In this early phase of reaction melts are largely viewed to be fragile (m > 20, Fig. 6a). It is for these reasons that in the first $t_R$ = 72h, changes in melt stoichiometry in the top (h = 9, $\Delta x$ = 5.5%) are nearly twice as large as in the bottom (h = 1, $\Delta x$ = 3%) of the column. However, as homogenization proceeds further it slows down qualitatively; note that melt stoichiometry in the top half is stuck near the super-strong melt composition x = 22% (in the shaded region) and takes almost 6 days to move from x = 22% to 23%, while in the lower half of the melt column it decreases from 24.5% to 23% in 5 days. In the inset of Fig. 3, we plot the variance $\sigma_x(t_R)$ of melt stoichiometry as a function of $t_R$, and find that it takes longer to homogenize melts at x =23% than at x = 19% or 21%.

The slow homogenization behavior above can be traced to the high viscosity of melts near the fragility minimum at x = 22% (Fig. 6a) at the high reaction temperature (950°C). Consider the Vogel Fulcher-Tammann plot of Fig.7 for two cases, a super strong melt, m = 15 and a fragile one at m = 30. A super-strong melt will, in general, display an Arrherian variation of shear relaxation time $\tau(T)$ or viscosity ($\eta(T)$) across the range of $T_g/T$. On the other hand, for the fragile melt m = 30, a bowing of the $\tau(T)$ occurs particularly in the middle of the range of $T_g/T \sim 0.5$. For the reaction temperature T = 950°C, and a $T_g$ of 200°C, the ratio $T_g/T = 0.39$. One thus expects viscosity ($\eta = G\tau$) of the super-strong melts to be about two orders of magnitude greater than those of fragile melts away from that minimum (Fig. 4). A perusal of the h(x) plots (Fig.4) shows that in the top half of the melt column the diffusion slows down *qualitatively* once melt stoichiometry approach the viscous *super-strong* compositions in the narrow dark blue band. In the lower-half of the tube (h < 5) diffusion is not directly hindered by the *super-strong* compositions, and the melt stoichiometry 'x', steadily reaches the end value of 23% as the system globally homogenizes to attain the weighed nominal composition, within less than a 1/4% error in x. A statistical analysis of the composition spread can be followed in time by computing the variance $\sigma^2_x(t) = \langle x(t)^2 \rangle - \langle \bar{x}(t) \rangle^2$, where averages are performed over the height of the tube. Results are shown in the inset of Fig. 3, and reveal that $\sigma_x(t)$ evolves quite differently with composition. At x=23% Ge, the estimated time of homogenization appears to be about 200 h, i.e., much longer than those for compositions lying outside the fragility minimum of super-strong melts (21% and 19%). Furthermore, the asymmetric homogenization at the fixed synthesis temperature (but at different $T_g(x)/T$) in the Se- or Ge-rich liquids cannot result from the difference in $T_g$. An estimate of the melt diffusivity at 950° C using both the Eyring equation[10] $D=k_BT/\eta$ and different empirical relationships for melt viscosity[4,26,27] determined from our measured m(x), shows that diffusivity decreases by a factor of $10^2$-$10^3$ between the x = 22% composition and those compositions lying outside the IP window (light blue band in Fig. 6a). Thus, small compositional variations along a batch result from diffusivity barriers that slow down the homogenization process qualitatively.

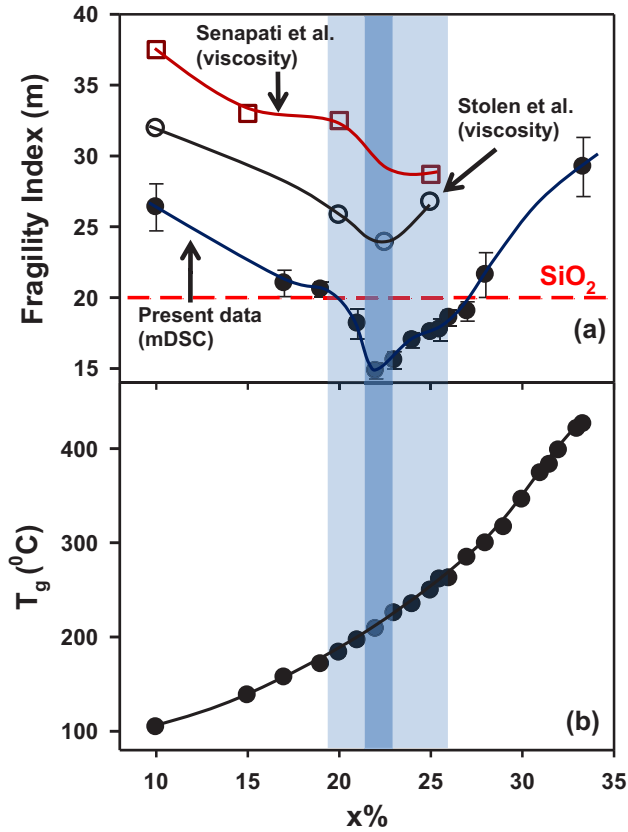

**Figure 6:** (a) Fragility index vs. x% reported by Senapati et al.[12] (□) and Stolen et al.[14] (○) using viscosity measurements. Fragility index of the present work (●) was obtained from Cp* measurements. (b) $T_g$ variation vs. x% reported by Bhosle et al.[19]

The significantly lower value of m observed in mDSC experiments at x = 22.5% in the specially homogenized melts than in viscosity measurements is curious but cannot be attributed to the methods given that both dielectric and calorimetric spectroscopies and viscosity measurements usually yield the same fragility[28]. It may be related to the use of much larger sized melts (>50 grams) of less homogeneity[29] in the viscosity measurements. The present value of the fragility index for the x = 22% Ge composition is lower than the celebrated example of silica[4] which, to date, is very close to the reported[9] theoretical lower limit of m (14.93), obtained from a topology derived equation for the viscosity change.

**B. Topology, fragility, reversibility window and the glass transition**.

More general correlations emerge from the present results between fragility, $T_g$, molar volumes and the IP that are linked through network topology, and we comment on these next. As has been discussed elsewhere [5,16] the IP represents a rigid but stress-free phase of these glasses that have some remarkable properties including their space filling nature, weak ageing (as compared to the flexible and stressed rigid phase), and presence of extended range structural correlations that lead to adaptation and isostatic character. The notion of adaptation under increasing stress (i.e. bond density), or self-organization that is central to the IP was first demonstrated from simple phenomenological models[30-32] and more recently from Molecular Dynamics Simulations [33,34].

The behavior of both the activation energy and the fragility with composition (Fig. 5 and 6), and the sharp minimum in the centroid of the IP suggests that there are strong underlying connections with thermodynamic signatures at the glass transition. This connection reminds us of the reported relationship between kinetic (m) and thermodynamic fragility (heat capacity jump $\Delta C_p$) as popularized by Angell[35]. In fact fragile liquids display a rapid change of structure with temperature leading to large changes in configurational entropy, and resulting in a large jump in the heat capacity, $\Delta C_p$, across $T_g$. This behavior should be contrasted to the one expected for strong liquids. The latter possessing strong directional bonds (covalent interactions modified by ionic ones) usually produce much more stable behavior in transport/thermodynamic properties with increasing temperature. Here, one has to recall that $\Delta C_p$ accessed from differential scanning calorimetry (DSC) measures not only vibrations, rotations, and translations but also enthalpic changes associated with relaxation at large length scales. On the other hand, modulated-DSC permits separating the endothermic heat flow near $T_g$ into thermal contributions (reversing heat flow) to $C_p$ from the kinetic ones (non-reversing heat flow). The latter capture most of the enthalpic relaxation associated with the slow-down of dynamics as a liquid approaches $T_g$.

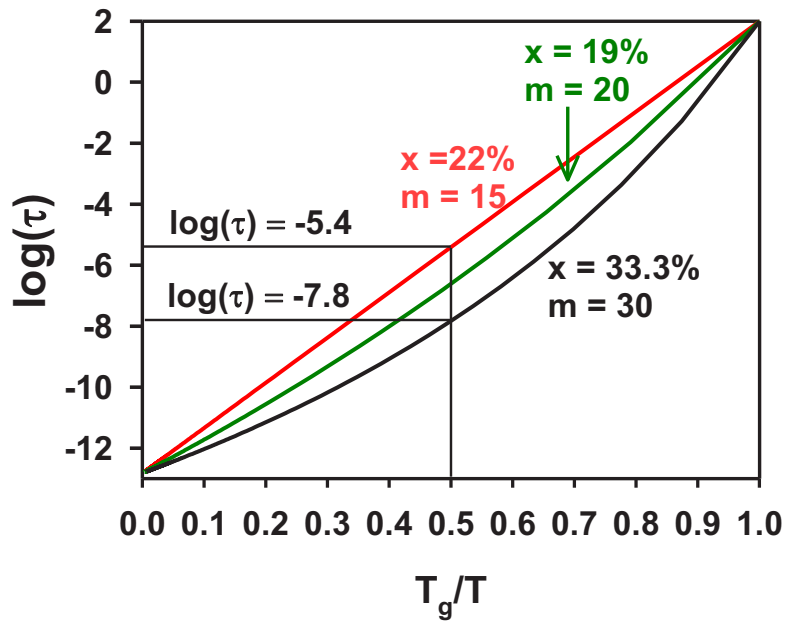

**Figure 7:** Logarithm of average relaxation time ($\tau$) plotted vs. inverse temperature normalized to Tg. Curves corresponding to different fragilities were calculated using Vogel-Fulcher equation rewritten as $\log(\tau) = \log(\tau_g) - m_{min} + m^2_{min}(T_g/T)/[m - (m-m_{min})T_g/T]^4$

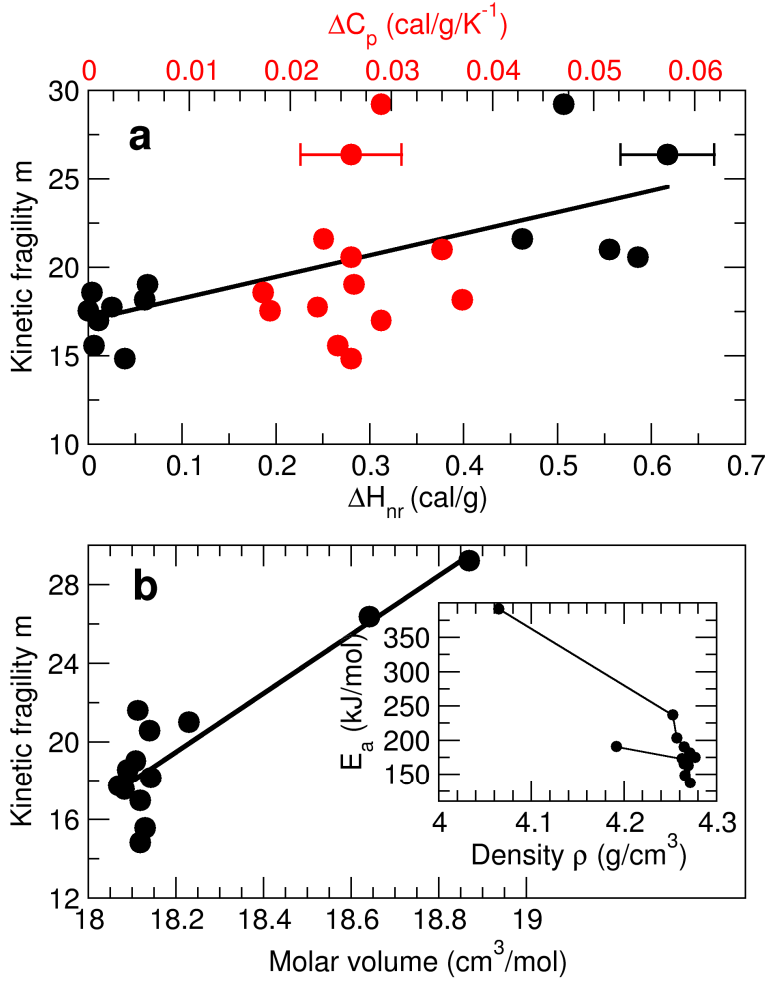

**Figure 8:** Kinetic fragility of Ge-Se melts as a function of (a) the non-reversing heat flow $\Delta H_{nr}$ and the heat capacity jump $\Delta C_p$ at the glass transition (red, upper axis), and (b) molar volume and density (inset). Error bars are about the same for all compositions.

In Fig. 8, we plot different correlations established from our experimental data on Ge-Se melts. Fig. 8(a) shows the behavior of the kinetic fragility m to increase as a function of the non-reversing heat flow $\Delta H_{nr}$, whereas no such correlation is observed between m and $\Delta C_p$. We are thus led to believe that contributions of vibrations to $C_p$ dominate at low temperatures (where $dH_{nr}/dt = 0$) while those of translations/rotations which grow as the liquid goes through $T_g$ lead to minor contributions to the total heat flow and fragility. The latter is obviously controlled by longer range relaxations which are embedded in the non-reversing heat flow, and thus m grows with $\Delta H_{nr}$ at a slope 12.2 g/cal (Fig. 8a).

Given the limited variation of m (29-14.8) accessed over the whole range of Ge-Se melt compositions examined, it is hard to conclude if these correlations would extend to all other glass-forming liquids. In the much more fragile organic liquids we are unaware of mDSC measurements using the decomposition of heat flow into reversing and non-reversing components except for a polychlorinated biphenyl (PBD, fragility index m=74.3), which shows $\Delta C_p$=0.06 cal/g/K and $\Delta H_{nr}$=0.813 cal/g for similar heating/cooling rates (3°C/min against 2°C/min, see[36]). This is a system with non-directional bonding, and apparently also shows the non-reversing heat flow to increase with fragility, but we also recognize that the heat capacity jump ($\Delta C_p$) also increases with fragility. Other examples of organic liquids (glycerol, salol, Propylene glycol) with increased fragility can only be compared qualitatively given that an analysis of the heat flow in terms of complex $C_p$ has been preferred[28]. Inspection of the imaginary part $C_p''(\omega)$ (related to the non-reversing heat flow) for organic liquids having large fragility (m~40-70) reveals that the term exceeds 70 mcal/g/°C at the peak or maximum value when $\omega\tau$=1. These $C_p''(\omega)$ terms are much larger than those observed by us in Ge-Se liquids (Fig. 4). We should note though that the cooling/rate used in $C_p''(\omega)$ measurements of 0.5 K/min are smaller than those used (3.0 K/min) in the non-reversing heat flow measurements, and these factors will obviously affect results of the measurements.

It is well known that fragility of glass-forming systems is strongly influenced by liquid density. While polymers are quite successfully described[37] in terms of free volume models[38,39] it is useful to explore if such an approach works for inorganic glass-formers. Previous results on the Ge-Se binary[18] reveal that molar volumes show a minimum near 22 mole% of Ge, giving an indirect indication that the fragility minimum may be correlated to the space-filling tendency (or density maximum) of glass compositions in the IP. Instead of plotting $E_a$ or m as a function of melt compositions as we have done earlier (Fig. 5 and 6), one can instead plot (Fig. 8b) m as a function of the glass density. This has the advantage that one can explore if fragility is correlated to molar volumes (Fig.8b). A visual inspection of the plot shows that the result reveals a generally known behavior, i.e., both temperature and density control dynamics. Indeed, fragility itself is a consequence of the relative interplay of temperature with density effects near $T_g$. Specifically, a strong behavior reflects a substantial contribution from density leading to jammed dynamics as becomes apparent from the limit in density (~4.28 g/cm$^3$, inset of Fig. 8) reached at the fragility minimum. On the contrary, relaxation of fragile liquids is more thermally-activated as demonstrated from a detailed investigation of many inorganic and polymeric glass-formers[40]. Given that flexible (x<20%) and stressed rigid (x>26%) melts are more fragile (as compared to IP compositions), corresponding energy barriers for relaxation must obviously be associated with (low energy) floppy modes and stress, respectively. Both nearly vanish in the IP, leading to the observed special relaxation behavior for compositions between 20 and 25% Ge.

### C. Revisiting Scaling of fragility with $T_g$ −some anomalies

Given the new results on the Ge-Se binary from present work, we investigate the validity of proposed scaling laws for fragility using the apparent activation energy for viscous relaxation[41]. Using the definition of m from equ. (1), and assuming a VFT of the form $\exp[A/(T - T_0)]$, one can actually calculate fragility and the apparent activation energy $E_g$ (the slope of the relaxation time at $T_g$) as:

$$m = \frac{AT_g}{(T_g-T_0)^2 \ln 10} \tag{2}$$

and:

$$E_g = \frac{AT_g^2}{(T_g-T_0)^2}. \tag{3}$$

As $T_g$ is of the same order of $T_0$, equations (2) and (3) above show that $E_g$ and m will scale respectively as $T_g^2$ and $T_g$. These results can be independently established from the Williams-Landel-Ferry (WLF) equation[42]. In WLF approach, the superposition parameter $a_T$ =$\eta T\rho(T)/\eta_g T_g \rho_g$, at the reference temperature $T_g$, can be written as:

$$\log a_T = \frac{C^g_1(T-T_g)}{C^g_2+T-T_g} \tag{4}$$

From equ. (4), fragility and apparent activation energy can be computed:

$$m = \frac{C^g_1}{C^g_2}T_g \tag{5}$$

$$E_g = \ln 10 \frac{C^g_1}{C^g_2}T_g^2 \tag{6}$$

Both VFT and WLF parameters being related through:

$$T_0 = T_g - C^g_2 \tag{7}$$

$$A = C^g_1 C^g_2 \ln 10 \tag{8}$$

Qin and McKenna[41] have shown from a compilation of experimental data that both scaling laws (5) and (6) are fulfilled in hydrogen bonded organics, polymeric and metallic glass formers, while inorganic glass formers appear to have their fragility nearly independent of $T_g$.

To test the validity of such scaling laws, we plot in Fig.9 the present results on the Ge-Se binary along with results for two other chalcogenide melts[43,44]. These results clearly indicate that the correlation established by Qin and McKenna holds for IP and stressed-rigid compositions in Ge-Se melts as seen from the linear increase of m with $T_g$ at x > 20% Ge, and from the linear increase of the activation energy $E_a$ with $T_g^2$ (inset of Fig. 9). A regression line for the Ge-Se data leads to m~-17.356+0.06 $T_g$ and $E_a$~-97.295+9.98$T_g^2$. The corresponding slope for the fragility index variation with Tg in the present inorganic melts (0.06) appears to be lower than those obtained[41] for polymers (0.28), metallic glass formers (0.17) and H-bonded liquids (0.25), but nevertheless the correlation is clearly visible. The latter correlation was not recognized in Qin and McKenna's compilation which included inorganic glass formers that were either stoichiometric or too sparse in composition to yield definite trends. Here, the systematic study on non-stoichiometric binary Ge-Se melts show trends similar to those established for network-forming polymeric liquids. In the case of the chalcogenide melts (As-Se and As-Ge-Se, Fig. 9), the correlation holds to a lesser extent in part because of smaller number of compositions studied, and possibly because of the inhomogeneity of melts examined (see [17]). The onset of nanoscale phase separation in the As-Se binary melts at higher As content (> 40 mole %), a second branch of the curve appears with a negative slope as shown in Fig. 9, as red squares.

Interestingly, an anti-correlation is detected in the flexible phase for the three families of chalcogenides, which cannot be inferred from equations (5)-(6). In fact, both m and $E_a$ are found to decrease with increasing network connectivity in these families of chalcogenides, that results in a continuous increase of the glass transition temperature $T_g$ as melt fragility m decrease with composition (Fig. 2). These flexible melts appear quite special given that the qualitative (positive) correlation between m and $T_g$ has been verified on a large number of glass-forming melts[45,46] and so has been the increase of fragility with cross-link density[47-49].

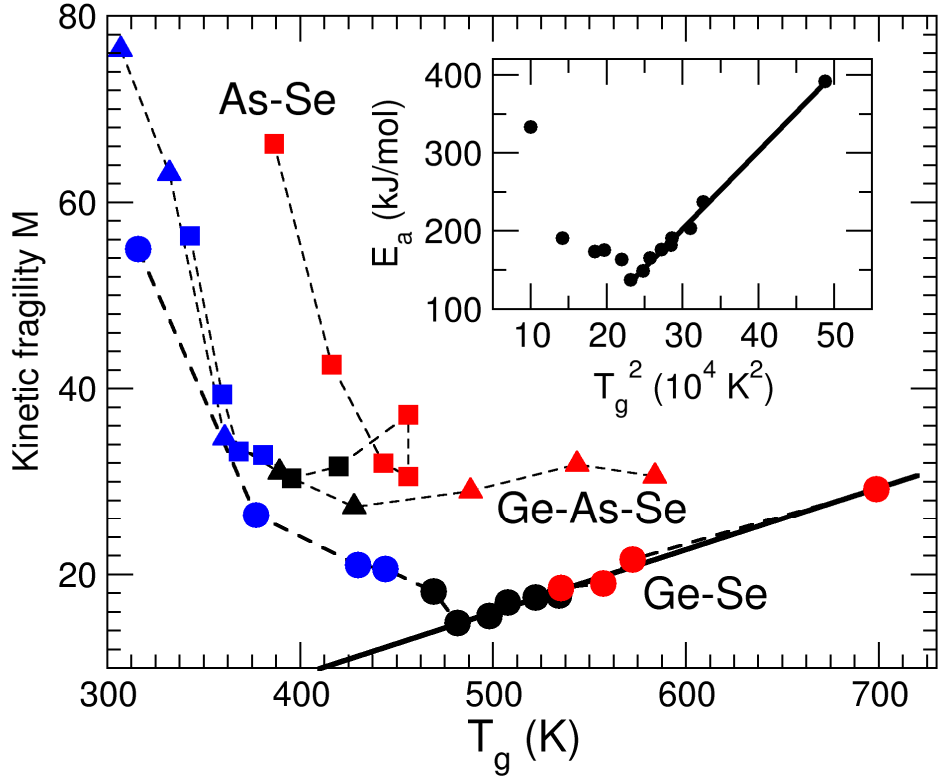

**Figure 9:** Fragility as a function of glass transition temperature in Ge-Se liquids (circles), together with previous data on As-Se (squares)[43,44], and As-Ge-Se (triangles)[13]. For each system, stressed rigid, IP and flexible compositions are marked in red, black and blue, respectively. The phase boundaries for As-Se and As-Ge-Se have been established from previous work[50,51]. The inset shows the activation energy $E_a$ as a function of $T_g^2$ for the present Ge-Se melts.

A negative slope in $m(T_g)$ and $E_a(T_g)$ would result if either $C^g{}_2 < 0$ or $C^g{}_1 < 0$. Both these conditions are improbable given that the former condition leads to $T_0 > T_g$ (divergence of the relaxation at a temperature higher than $T_g$) and the latter to $A < 0$ (decrease of the relaxation time with decreasing temperature). The unexpected behavior is even more striking given that WLF equations usually work very well for polymeric glass-forming liquids[52-54] and flexible Ge-Se melts actually fall in this category of materials. They are composed of polymeric Se chains that are weakly cross-linked by Ge atoms. Furthermore, the glass transition variation of such chalcogen-rich glasses are found to be accurately predicted by the Gibbs-Di Marzio equation[55,56] that was derived for cross-linked polymers.

At present we have no specific answer to this fragility anomaly of flexible melts in the Ge-Se binary reported in Fig. 9. Conventional wisdom suggests that all glass formers tend to possess an activation energy that increases with Tg[41]. However, with increasing composition, all chalcogenide glasses/melts lose their polymeric character to become fully 3D connected as their networks stiffen with increased Ge cross-links. The road toward the observed rigidity transition ultimately drives enthalpic, structural and volumetric changes, which in turn drive changes in dynamic properties.

### IV. Summary and conclusions

Physics of network glasses, as elucidated by Rigidity Theory, has stimulated[5,12,16-19,57-61] compositional studies of physical properties of melts and glasses. The observed fragility minimum thus appears to be intimately related to flexible to rigid transitions and the intermediate phase in corresponding glasses. An issue of central importance is how homogeneous must melts/glasses be in such studies to observe the intrinsic behavior of these thresholds? We believe compositional width of the percolative elastic phase transitions (stress and rigidity) provides a convenient scale. An estimate of width comes from the reversibility window wall, which we estimate[18] to be at $\Delta \bar{r} < 0.01$. Here $\bar{r} = 2(1 + x)$, designates the mean coordination number of the $Ge_xSe_{100-x}$ network taking Ge and Se to be 4- and 2- fold coordinated. The condition $\Delta \bar{r} < 0.01$, translates into a Ge stoichiometry variation $\Delta x < 1/2$ at.% across a melt composition, and thus fixes a measure of system homogeneity at a given x. For batch sizes greater than 2 grams, more care is needed to homogenize them. The variance in physical properties of chalcogenide glasses, such as for example, molar volumes of $Ge_xSe_{100-x}$ glasses reported by different groups[18] are much too large to be statistical, and reflect, in our view, the result of glass sample purity and heterogeneity.

In this work we have shown that certain liquids homogenized at the micron scale, super-strong behavior is manifested with a fragility index (m=14.8(0.5) even smaller than silica. Melts formed in the 21.5% < x < 23.0% range of Ge serve as a bottleneck to homogenization of $Ge_xSe_{100-x}$ melts/glasses. The narrow composition range resides near the center of the Intermediate phase[18,19] in corresponding glasses (Fig.6). The low value of m suggests existence of extended range structural correlations, microscopic reversibility, lack of network stress in such melts as in corresponding glasses, and provides a new perspective linking "strong melts" with network adaptability of that phase.

Finally, we observe that correlations between melt properties such as fragility or activation energy and thermal properties of glasses, such as $\Delta C_p$ and $\Delta H_{nr}$ that characterize $T_g$ can be established as highlighted by the present work. The observed linear relationship between fragility and the glass transition temperature is found to be fulfilled in the present Ge-Se melts. Nevertheless, we also find that flexible melts do not follow such correlations at all, and display instead a decrease of m with increasing glass transition temperature. The precise origin of this unexpected negative correlation remains to be established but it clearly has connections with the onset of rigidity. At this stage however, we can anticipate that the observed anomaly will be manifested generally. In network glass-forming melts, an increase (decrease) in connectivity leads always to a monotonic increase (decrease) of $T_g$[55]. This connectivity change drives a system from a flexible to a stressed rigid phase, and leads to a fragility minimum associated with an intermediate phase. As a consequence, the representation m(T$_g$) will always exhibit a minimum, and the usual positive slope for stressed rigid compositions.

This work is supported by NSF grant DMR-08-53957 and ANR grant No. 09-BLAN-0109-01.